
\magnification \magstep1
\raggedbottom
\openup 4\jot
\voffset6truemm
\headline={\ifnum\pageno=1\hfill\else
\hfill{\it One-loop effective potential for
$SO(10)$ GUT theories in de Sitter space}\hfill
\fi}
\def\cstok#1{\leavevmode\thinspace\hbox{\vrule\vtop{\vbox{\hrule\kern1pt
\hbox{\vphantom{\tt/}\thinspace{\tt#1}\thinspace}}
\kern1pt\hrule}\vrule}\thinspace}
\centerline {\bf ONE-LOOP EFFECTIVE POTENTIAL FOR}
\centerline {\bf SO(10) GUT THEORIES}
\centerline {\bf IN DE SITTER SPACE}
\vskip 1cm
\centerline {\bf Giampiero Esposito$^{1,2}$,
Gennaro Miele$^{1,3,*}$, Luigi Rosa$^{1,2}$}
\vskip 1cm
\centerline {\it ${ }^{1}$Istituto Nazionale di Fisica Nucleare}
\centerline {\it Mostra d'Oltremare Padiglione 20}
\centerline {\it 80125 Napoli, Italy;}
\centerline {\it ${ }^{2}$Dipartimento di Scienze Fisiche}
\centerline {\it Mostra d'Oltremare Padiglione 19}
\centerline {\it 80125 Napoli, Italy;}
\centerline {\it ${ }^{3}$NASA/Fermilab Astrophysics Center}
\centerline {\it Fermilab National Accelerator Laboratory}
\centerline {\it Box 500, Batavia, IL 60510-0500}
\centerline {\it United States of America.}
\vskip 3cm
\noindent
${ }^{*}$On leave of absence from the Dipartimento di Scienze Fisiche,
Universit\`a di Napoli, Italy.
\vskip 100cm
\noindent
{\bf Abstract.} Zeta-function regularization is applied to
evaluate the one-loop effective potential for
$SO(10)$ grand-unified theories in de Sitter
cosmologies. When the Higgs scalar field belongs to the
210-dimensional irreducible representation of $SO(10)$,
attention is focused on the mass matrix relevant
for the $SU(3) \otimes SU(2) \otimes U(1)$ symmetry-breaking
direction, to agree with low-energy phenomenology of the
particle-physics standard model. The analysis is restricted
to those values of the tree-level-potential parameters for
which the absolute minima of the classical potential have
been evaluated. As shown in the recent literature, such
minima turn out to be $SO(6) \otimes SO(4)$- or
$SU(3) \otimes SU(2) \otimes SU(2) \otimes U(1)$-invariant.
Electroweak phenomenology is more naturally derived, however,
from the former minima. Hence the values of the parameters
leading to the alternative set of minima have been
discarded. Within this framework,
flat-space limit and general
form of the one-loop effective potential are studied in detail
by using analytic and numerical
methods. It turns out that, as far as the
absolute-minimum direction is concerned, the flat-space limit
of the one-loop calculation about a de Sitter background does
not change the results previously
obtained in the literature, where the
tree-level potential in flat space-time was studied.
Moreover, when curvature effects are no longer negligible
in the one-loop potential, it is found that the early universe
remains bound to reach only
the $SO(6) \otimes SO(4)$ absolute minimum.
\vskip 4cm
\leftline {PACS numbers: 0260, 0370, 0420, 1115, 9880}
\vskip 100cm
\leftline {\bf 1. Introduction}
\vskip 0.3cm
\noindent
Over the last ten years, the idea by Coleman and Weinberg (1973)
on radiative corrections at the origin of spontaneous symmetry
breaking has played a very important role also in cosmology.
In particular,
in Allen (1983) and Allen (1985) the one-loop approximation of
path integrals in curved space was applied to study massless
scalar electrodynamics and $SU(5)$ non-Abelian gauge fields in
de Sitter space. For this purpose, the author used zeta-function
regularization (Hawking 1977, Esposito 1994), and was able to show
that the inflationary universe can only slide into either the
$SU(3) \otimes SU(2) \otimes U(1)$ or the $SU(4) \otimes U(1)$
extremum, in the case of $SU(5)$ gauge models. In his analysis,
Allen (1983, 1985) was dealing with Wick-rotated path integrals,
leading to a Riemannian background 4-geometry with $S^4$
topology and constant scalar curvature,
i.e. the Euclidean-time version of de Sitter space-time.

More recently, work by the authors (Buccella {\it et al}
1992, Esposito {\it et al} 1993) has led to
a deeper understanding of the results in Allen (1985).
However, since the
technique described in Allen (1985) enables one to evaluate
the one-loop effective potential for {\it all} non-Abelian
gauge theories in de Sitter space, a naturally occurring
question is whether one can repeat this analysis in the case
of physically more relevant GUT theories in de Sitter cosmologies.
For this purpose, our paper studies the one-loop effective
potential of $SO(10)$ GUT theories.

$SO(10)$ gauge theories as unified
models for strong, electromagnetic and weak
interactions (Fritzsch and Minkowski 1975, Tuan 1992)
have been studied over many years for their interesting physical
properties. There are several motivations for this choice,
and the strongest one can be found in the
predictions for nucleon lifetimes. In this case in fact, $SO(10)$
models enable one to obtain higher values for the masses of the
lepto-quarks which mediate proton decay and which were predicted to be too
low, with respect to the experimental
lower limit, in the minimal $SU(5)$ model.
This property is essentially related to the presence of an
intermediate symmetric phase between the $SO(10)$ symmetry at GUT scale
and the $SU(3)\otimes SU(2)\otimes
U(1)$ symmetry at weak scale.

A complete analysis of the possible symmetry-breaking
patterns with Higgs particles in
representations with dimension $\leq 210$ and with only an
intermediate symmetry group $G'$ between
$G$ and $SO(10)$ has led to only four
different possibilities for a physically relevant $SO(10)$ unified model
(Buccella 1988) [table I]. With the notation of table I,
$SU(3)_C$ is the colour group,
$SU(2)_{L,R}$ are the {\it left} and {\it right}
$SU(2)$ groups whose representations differ by their
behaviour with respect to helicity.
Moreover, $B-L$ is the difference between
baryonic and leptonic number, $D$ is the discrete
left-right interchanging symmetry
(Kuzmin and Shaposhnikov 1980, Chang {\it et al} 1984), and
$SU(4)_{PS}$ denotes the $SU(4)$
Pati-Salam group (Pati and Salam 1973).
For these models, using the one-loop
approximation for the renormalization-group
equations, the upper limit for
the values of the symmetry-breaking
scales of $SO(10)$ ($M_X$), and of $G'$
($M_R$), is reported for the different
models in their minimal formulation [table II].

As one can see, both models without
$D$ symmetry yield sufficiently high
values for the scale $M_X$, and the model
with $G'\supset SU(4)_{PS}$ predicts
$M_{R}=10^{11}$ GeV, while the one with
$G'\supset SU(3)_{C}\otimes U(1)_{B-L}$ gives
rise to a value about two orders of magnitude smaller.
Using these results and their implications
for proton decay we can safely
restrict our analysis of the cosmological
implications of $SO(10)$ GUT models, to
the ones which appear physically more
relevant and which contain the Higgs field in the
210-dimensional irreducible representation.

Our paper is organized as follows. Section 2, aimed at
cosmologists who are not familiar with
grand-unified theories, describes
the basic elements of $SO(10)$ GUT models in particle physics,
the tree-level potential for the 210-dimensional irreducible
representation, and the mass matrix relevant for the
$SU(3) \otimes SU(2) \otimes U(1)$ symmetry-breaking direction.
Section 3 derives the one-loop form
of the effective potential for $SO(10)$ GUT theories studied
about a de Sitter background. The numerical analysis of the
corresponding flat-space limit is then carried out in section 4.
Section 5 studies by numerical methods the one-loop effective
potential in the region where no asymptotic expansion
for infinite or vanishing curvature
of the special function occurring in such a potential
can be made. Results and
concluding remarks are presented in section 6.
\vskip 0.3cm
\leftline {\bf 2. SO(10) GUT theory in flat space-time}
\vskip 0.3cm
\noindent
The group $SO(10)$ is defined as the set of $10\times10$
orthogonal matrices with unit determinant, and with the usual product
rules. It has $45$ generators here
denoted by $T_{ij}$ ($i,j=0,1,\ldots,9$) obeying the following
commutation relations:
$$
\Bigr[T_{jk},T_{lm}\Bigr]=i
\Bigr(\delta_{jl}~T_{mk}+\delta_{jm}~T_{kl}+\delta_{kl}~T_{jm}+
\delta_{km}~T_{lj}\Bigr).
\eqno (2.1)
$$
Considering the vector irreducible representation
($\underline{10}$) of the group,
which we indicate by $\varphi_{l}$, the action of the
generators $T_{jk}$ on it is given by
$$
T_{jk}~\varphi_{l} \equiv
i\Bigr(\delta_{kl}~\varphi_{j}-\delta_{jl}~\varphi_{k}\Bigr).
\eqno (2.2)
$$

To construct a satisfactory gauge
theory based on the $SO(10)$ local symmetry,
which has the proper residual symmetry
in the low-energy limit, we need a
Higgs mechanism to break the symmetry
spontaneously (O'Raifeartaigh 1986). This is based on the presence
of a fundamental scalar particle
(Higgs field), belonging to one or more
irreducible representations (hereafter referred to as
IRR's) of the gauge group, whose
dynamics is ruled by a Higgs potential.
In the present case, we are going to
study the most general, renormalizable
and conformally invariant Higgs
potential constructed by using only the
IRR $\underline{210}$, which is obtained by
the completely anti-symmetrized product
of four different $\underline{10}$'s as
$$
\Phi_{abcd}=N \; \mu_{[a}\otimes
\nu_{b}\otimes\rho_{c}\otimes
\sigma_{d]}
\eqno (2.3)
$$
where $N$ is the normalization constant.
The $\underline{210}$ IRR has four independent quartic
invariants, i.e. ${\| \phi \|}^{4}$ and three non-trivial
invariants (see (2.8)-(2.10)),
hence the Higgs potential we are going to construct will be a
function of these.
Multiplying two $\underline{210}$ and symmetrizing one gets
the Clebsch-Gordan decomposition
$$
\left(\underline{210}\otimes\underline{210}\right)_{sym.}=
\underline{1}\oplus\underline{45}\oplus\underline{54}\oplus
\underline{210}\oplus\underline{770}\oplus(\underline{1050}
\oplus\overline{\underline{1050}})\oplus\underline{4125}
\oplus\underline{8910}\oplus\underline{5940}
\; \;
\eqno (2.4)
$$
where $\underline{45}, \underline{54}, \underline{770}$ and so on
denote the IRR's with dimension 45, 54, 770 respectively.

The IRR's $\underline{45}$, $\underline{210}$ and $(\underline{1050}\oplus
\overline{\underline{1050}})$ give no contribution along the
$SO(6) \otimes SO(4)$-invariant direction.
This can be easily understood noticing that
$\underline{45}$ and $(\underline{1050} \oplus \overline{\underline{1050}})$
do not contain singlets along the above direction and that the only singlet
contained in the $\underline{210}$ representation is such that
$
C^{~~\underline{210}~~
{}~~\underline{210}~~~~\underline{210}}_{(1,1,1)~
(1,1,1)~(1,1,1)}=0
$.
With our notation, $(1,1,1)$ is the only singlet with
respect to the $SU(4) \otimes SU(2) \otimes SU(2)$ group contained in
the $\underline{210}$ representation. Moreover, we study
the Clebsch-Gordan coefficient (Cornwell 1984b)
corresponding to the decomposition $(1,1,1)_{\underline{210}}\otimes
(1,1,1)_{\underline{210}} \rightarrow (1,1,1)_{\underline{210}}$.

Hence the only quartic non-trivial invariants, apart from the
fourth power of the $\underline{210}$ norm
which is isotropic in the space of IRR's and
hence cannot discriminate among the invariant directions, are
$\|(\phi \phi)_{\underline {45}}\|^2$,
$\|(\phi \phi)_{\underline {210}}\|^2$ and
$\|(\phi \phi)_{\underline {1050}}\|^2$
(where for example, the symbol
$(\phi \phi)_{\underline {45}}$
stands for the $\underline{45}$
IRR contained in the product of two $\underline{210}$).

By virtue of the above considerations, the
most general renormalizable and conformally
invariant Higgs potential, made out of the
$\underline{210}$ representation only, turns out to be
a linear combination of the above invariants, with arbitrary coefficients
$g_{1}$, $g_{2}$, $g_{3}$ and $\lambda$
$$
V(\phi) = g_{1}~\|(\phi \phi)_{\underline {45}}\|^{2}
+ g_{2}~\|(\phi \phi)_{\underline {210}}\|^{2}+
g_{3}~\|(\phi \phi)_{\underline {1050}}\|^{2}
+\lambda~ \|\phi\|^{4}.
\eqno (2.5)
$$

The IRR $\underline{1050}$ is quite complicated. We thus
prefer to express the term
$\|(\phi \phi)_{\underline {1050}}\|^{2}$ as a function of
the representations $\underline{45}$, $\underline{54}$ and $\underline{210}$
$$
\|(\phi \phi)_{\underline {1050}}\|^{2}
= - {35 \over 6} \|(\phi \phi)_{\underline {45}}\|^{2}
- {7 \over 3}\|(\phi \phi)_{\underline {54}}\|^{2}
+ {5 \over 4} \|(\phi \phi)_{\underline {210}}\|^{2} +
{1 \over 10} \|\phi\|^{4}.
\eqno (2.6)
$$
In other words, since the space of group invariants is a vector
space, we can evaluate the components of
$\|(\phi \phi)_{\underline {1050}}\|^{2}$ along the basis vectors.
Thus, by inserting (2.6) in (2.5)
we get the flat-space potential
$$ \eqalignno{
V(\phi)&=\left(g_{1}-{35 \over 6} g_{3} \right)
{}~\|(\phi \phi)_{\underline {45}}\|^{2}
+\left(g_{2} + {5 \over 4} g_{3} \right)
{}~\|(\phi \phi)_{\underline {210}}\|^{2}\cr
&-{7 \over 3} g_{3}~\|(\phi \phi)_{\underline {54}}\|^{2}
+\left({1 \over 10} g_{3}
+\lambda \right)~ \|\phi\|^{4} .
&(2.7)\cr}
$$

To clarify the definitions of $(\phi \phi)_{\underline {45}}$,
$(\phi \phi)_{\underline {210}}$
and $(\phi \phi)_{\underline {54}}$,
we point out that from the symmetrized product of two
IRR's $\underline{210}$ ($\phi_{abcd}$) it is possible,
using the Levi-Civita symbol $\epsilon_{i_{0}...i_{9}}$, to
construct the IRR (hereafter we sum over repeated indices)
$$
(\underline{45})_{ab} = C^{\underline{210}~~
\underline{210}~~\underline{45}}_{cdef~ghil~ab}~\phi_{cdef}~
\phi_{ghil}
={1 \over \sqrt{70}}\epsilon_{abcdefghil}~\phi_{cdef}~\phi_{ghil}.
\eqno (2.8)
$$
Analogously, using the $SO(10)$ invariance of the Levi-Civita
symbol, the $\underline{210}$ representation can be denoted
by $4$, or, equivalently, $6$ indices of the completely
antisymmetric tensor
$$ \eqalignno{
(\underline{210})_{abcd}&=(\underline{210})_{efghil}
= C^{\underline{210}~~~\underline{210}~~~~
\underline{210}}_{abmn~mncd~efghil}
{}~\phi_{abmn}~\phi_{cdmn}\cr
&={1 \over \sqrt{90}}\epsilon_{abcdefghil}~\phi_{abmn}~\phi_{cdmn}
{}~~~
&(2.9)\cr}
$$
and
$$
(\underline{54})_{ab} = {1 \over \sqrt{112}} \left( \phi_{amno}~
\phi_{bmno}~+~\phi_{bmno}~\phi_{amno}\right)~~~~~~a \neq b .
\eqno (2.10)
$$
If $a=b$ we have $9$ more terms orthogonal to the trace, here
omitted for the sake of brevity.

Starting from the general potential (2.7), a complete analysis of its
absolute minima would require first of all, the computation of the above
potential along every direction of possible residual symmetry and secondly
the determination of the ranges for the parameters $g_{i}$ corresponding
to the different residual symmetries for the absolute minima. This is exactly
what was done in the case of $SU(5)$ (Allen 1985, Buccella {\it et al} 1992,
Esposito {\it et al} 1993) with the Higgs scalar field in the
adjoint representation.

Unfortunately, the technical difficulties due to
the complexity of the group $SO(10)$ with respect to the unitary
ones and the size of the IRR used,
make it impossible to extend the previous analysis to
the present case. For this reason, at least at this stage, we
restrict our considerations to the study of the modifications, induced
by one-loop and curvature effects (section 3),
of the symmetry-breaking pattern,
for choices of the parameters $g_{i}$ corresponding to absolute minima
of the potential at tree-level, invariant under the residual-symmetry
group $SU(3) \otimes SU(2) \otimes U(1)$. These are the only ones relevant
for particle physics in flat space, because they predict
the correct low-energy-limit phenomenology.

The most general singlet $\phi_{0}$ with respect to the group
$SU(3) \otimes SU(2) \otimes U(1)$ contained in the $\underline{210}$
representation is
$$ \eqalignno{
\phi_{0}&= {z_{1}\over \sqrt{3}} \Bigr(\phi_{1234}+\phi_{3456}+
\phi_{5612}\Bigr)\cr
&+{z_{2} \over \sqrt{6}}\Bigr( \phi_{1278}+\phi_{3478}+\phi_{5678}+
\phi_{1290}+\phi_{3490}+\phi_{5690} \Bigr) + z_{3}\phi_{7890}
&(2.11)\cr}
$$
where $\Bigr(z_{1}^{2}+z_{2}^{2}+z_{3}^{2}\Bigr)=1$.
Varying the $z_{i}$ parameters
in their ranges, we get the following residual-symmetry groups
(see comments in section 1):
$$
z_{2}=0 \rightarrow  SU(3)_{C} \otimes SU(2)_{L} \otimes SU(2)_{R}
\otimes U(1)_{B-L}
\eqno (2.12a)
$$
$$
z_{2}=z_{3}=0 \rightarrow SU(3)_{C} \otimes SU(2)_{L}
\otimes SU(2)_{R} \otimes U(1)_{B-L} \times D
\eqno (2.12b)
$$
$$
z_{1}=z_{2}=0 \rightarrow SU(4)_{PS} \otimes SU(2)_{L} \otimes
SU(2)_{R}
\eqno (2.12c)
$$
$$
{z_{1} \over \sqrt{3}} = {z_{2} \over \sqrt{6}} = z_{3}
\rightarrow SU(5) \otimes U(1)
\eqno (2.12d)
$$
$$
{\rm otherwise} \rightarrow SU(3)_{C} \otimes SU(2)_{L}
\otimes U(1)_{T_{3R}} \otimes U(1)_{B-L}
\; \; \; \;
\eqno (2.12e)
$$
where $T_{3R}$ is the $z$-component of the $SU(2)_R$ group.

Inserting (2.11) in (2.7) one gets in flat space-time
$$
{\widehat V} \equiv V(\phi_{0}) = \biggr({\alpha \over 8}f_{\alpha}
+{\gamma \over 4}f_{\gamma}+{\delta \over 9}
f_{\delta} + (\lambda - \delta) \biggr)
\| \phi_{0} \|^{4}
\eqno (2.13)
$$
where
$$
\alpha \equiv {4\over 945}\Bigr(-108g_{1}+28g_{2}+140g_{3}\Bigr)
\eqno (2.14)
$$
$$
\gamma \equiv {8\over 35}g_{1}
\eqno (2.15)
$$
$$
\delta \equiv -{1\over 10}g_{3}
\eqno (2.16)
$$
$$
f_{\alpha} \equiv {\Bigr(z_{1}^{2}+z_{2}^{2}\Bigr)}^{2}
+z_{2}^{2}{\Bigr(2z_{1}+\sqrt{3}z_{3}\Bigr)}^{2}
+{3\over 4}z_{2}^{4}
\eqno (2.17)
$$
$$
f_{\gamma} \equiv {\biggr(z_{1}z_{3}+{z_{2}^{2}\over \sqrt{3}}
\biggr)}^{2}+(z_{1}z_{2})^{2}+f_{\alpha}
\eqno (2.18)
$$
$$ \eqalignno{
f_{\delta}& \equiv 30{\biggr(z_{1}z_{3}+{z_{2}^{2}\over \sqrt{3}}
\biggr)}^{2}
+30z_{1}^{2}z_{2}^{2}
+{\biggr(2z_{1}^{2}-{z_{2}^{2}\over 2}-3z_{3}^{2}\biggr)}^{2}\cr
&+5{\Bigr(z_{1}^{2}+z_{2}^{2}\Bigr)}^{2}
+5z_{2}^{2}{\Bigr(2z_{1}+\sqrt{3}z_{3}\Bigr)}^{2}
+{15\over 4}z_{2}^{4}.
&(2.19)\cr}
$$
Since in the following analysis
$\delta$ is always negative and $\alpha$ may take
negative values, the tree-level potential (2.13) is unbounded
from below, unless we impose the restriction
$$
\lambda \geq {\mid \alpha \mid \over 8}
\Bigr(f_{\alpha}\Bigr)_{{\rm max}}
+{\mid \delta \mid \over 9}
\Bigr(f_{\delta}\Bigr)_{{\rm max}}.
\eqno (2.20)
$$
Note also that contributions
proportional to a cubic term in the potential
(denoted by $\beta$ in Acampora {\it et al} (1994)) are set to zero,
since we are assuming conformal invariance of our model
(Buccella {\it et al} 1992, Esposito {\it et al} 1993).
This assumption enables one to be more predictive, because it leads
to a smaller number of free parameters.
In the models proposed in Acampora {\it et al} (1994) a complete study of
the potential at tree-level for the case $z_{2}=0$ has been carried out,
including the range of the bare-potential parameters such
that the absolute minimum lies in the two-dimensional surface ($z_{2}=0$).

However, since we are interested in the modification of the bare potential
produced at one-loop by de Sitter curvature,
we can use only part of the inequalities appearing in Acampora
{\it et al} (1994).
More precisely, the parameters are bound to lie in regions where the mass
spectrum is positive and the first derivatives of the effective potential
vanish.

Thus, the allowed ranges for the parameters become

(1) $z_{1}=0~~,~~z_{3}=1~\Rightarrow~SO(6) \otimes SO(4) \sim SU(4) \otimes
SU(2)_{L} \otimes SU(2)_{R}$
$$
\gamma>0~~ \delta<0 ~~ \beta=0 ~~ \alpha> -2\gamma
$$

(2) $z_{1}^{2} + z_{3}^{2}=1~~\Rightarrow~
SU(3)_{C} \otimes SU(2)_{L} \otimes
SU(2)_{R} \otimes U(1)_{B-L}$
$$
\alpha>0~~ \beta=0 ~~ -{3 \over 5}\alpha< \gamma <- {1\over 2} \alpha
$$
$$
{3\Bigr(9\alpha^{2}+9\alpha \gamma -18\gamma^{2}
+4\Bigr(3\alpha+7\gamma\Bigr)\sqrt{-3\gamma(\alpha+\gamma)}\Bigr)
\over
320 \Bigr(\gamma-\sqrt{-3\gamma(\alpha+\gamma)}\Bigr)}
< \delta <
{3\gamma^2 \over 10 (3\alpha + \gamma)}  .
$$
In the case $z_{1}=1,z_{3}=0$ one finds that it is
impossible to get positive mass for both (6,2,2,-2/3) and (1,2,2,2).
In fact this is a saddle point in the space representation.
Indeed, since we are interested in the case when the intermediate
symmetry group contains $SU(4)_{PS}$ for the reasons described
in the introduction, we can restrict our analysis to case (1).

For our purposes we need to compute the mass matrix for
the gauge bosons. This comes from the kinetic term
for the Higgs field when we expand this scalar field around its vacuum
expectation value $\phi_{0}$ to get
$
\Bigr(D_{\mu} \phi_{0}, D^{\mu} \phi_{0} \Bigr)
= {\cal G}^{2}
\Bigr(T_{ab} \phi_{0} , T_{cd} \phi_{0} \Bigr)
A^{ab}_{\mu} A^{cd \; \mu}
$,
where square brackets denote the
scalar product in the $210$-dimensional
space and ${\cal G}$ is the gauge coupling
constant of the $SO(10)$ group.
Since the general form of $\phi_{0}$ is given, we can evaluate
the action of the $45$ generators $T_{ab}$ on it.

One now takes the decomposition of the
adjoint representation $\underline{45}$ under the group
$SU(3) \otimes SU(2) \otimes U(1)$. This makes it necessary
to use a standard notation in particle physics, where $(l,r,x)$
denotes the tensor which behaves as an $l$-dimensional
representation under $SU(3)$, $r$-dimensional under $SU(2)$
and takes a value=$x$ when acted upon by the $U(1)$ generator.
By virtue of the Wigner-Eckart theorem (Cornwell 1984a),
defining $m^{2} \equiv {\cal G}^{2}
{\| \phi_{0} \|}^{2}$, and evaluating the Clebsch-Gordan
coefficients, one finds that the
non-vanishing eigenvalues $m^2_{(l,r,x)}$ are
$m^2_{(1,1,1)}=m^2_{(1,1,-1)}=m^2 \left[{z_{2}^{2}\over 2}\right]$ with
degeneracy $1$, $m^2_{(3,1,2/3)}=
m^2 \left[{2\over 3}\Bigr(z_{1}^{2}+z_{2}^{2}\Bigr)\right]$
with degeneracy $3$, $m^2_{(3,2,1/6)}=
m^2 \biggr[{2\over 3}z_{1}^{2}+{z_{2}^{2}\over 2}
+z_{3}^{2}-\sqrt{{2\over 3}} \; z_{2}z_{3}\biggr]$
with degeneracy $6$, and
$m^2_{(3,2,-5/6)}=m^2 \biggr[{2\over 3}z_{1}^{2}+{z_{2}^{2}\over 2}
+z_{3}^{2}+{2\sqrt{2}\over 3}z_{1}z_{2}
+\sqrt{{2\over 3}} \; z_{2}z_{3}\biggr]$
with degeneracy $6$ as well. Note that
this is the mass matrix relevant for the
$SU(3) \otimes SU(2) \otimes U(1)$ symmetry-breaking direction.
This choice is motivated by low-energy phenomenology of the
particle-physics standard model, and all groups containing
$SU(3) \otimes SU(2) \otimes U(1)$ lead to the same kind of
mass matrix (of course, the $z_{i}$ parameters take different
values for different groups). Hence we only rely on the
$\phi_{0}$ singlet appearing in (2.11).
\vskip 0.3cm
\leftline {\bf 3. One-loop effective potential in de Sitter space}
\vskip 0.3cm
\noindent
Within the framework of inflationary cosmology, the quantization
of non-Abelian gauge fields has been recently studied in the
case of $SU(5)$ GUT theories (Allen 1985, Buccella {\it et al}
1992, Esposito {\it et al} 1993). In this case one starts from
a bare, Euclidean-time Lagrangian
$$
L={1\over 4}{\rm Tr}\Bigr({\bf F}_{\mu \nu}
{\bf F}^{\mu \nu}\Bigr)
+{1\over 2}{\rm Tr}\Bigr(D_{\mu}\varphi\Bigr)
\Bigr(D^{\mu}\varphi\Bigr)
+V_{0}(\varphi)
\eqno (3.1)
$$
where both the gauge-potential $A^{\mu}$ and the Higgs scalar
field $\varphi$ are in the adjoint representation of
$SU(5)$. Note that boldface characters are used to denote
the curvature 2-form ${\bf F}$ in the non-Abelian case, to avoid
confusion with the curvature 2-form $F$ in the Abelian case.
The background 4-geometry is de Sitter space with
$S^{4}$ topology. The background-field method is then used,
jointly with the gauge-averaging term first proposed by
't Hooft
$$
L_{g}={{\widetilde \alpha} \over 2}{\rm Tr}
{\Bigr(\nabla_{\mu}A^{\mu}-i{\cal G} {\widetilde \alpha}^{-1}
[\varphi_{0},\varphi]\Bigr)}^{2}.
\eqno (3.2)
$$
This particular choice is necessary to eliminate in the
total action cross-terms involving
${\rm Tr}\Bigr(\nabla_{\mu}A^{\mu}\Bigr)$ and the
{\it commutator} $[\varphi_{0},\varphi]$, where
$\varphi_{0}$ is a constant background Higgs field. After sending
${\widetilde \alpha} \rightarrow \infty$
(Landau condition), and denoting by $\Omega={8\over 3}\pi^{2}
a^{4}$ the volume of a 4-sphere of radius $a$,
the resulting one-loop effective potential is (Allen 1985)
$$ \eqalignno{
V(\varphi_{0})&=V_{0}(\varphi_{0})
+{1\over 2\Omega}\log \; {\rm det} \; \mu^{-2}
\Bigr[\delta_{ab}\Bigr(-g_{\mu \nu}\cstok{\ }
+R_{\mu \nu}\Bigr)+g_{\mu \nu}M_{ab}^{2}
(\varphi_{0})\Bigr]\cr
&+{1\over 2\Omega}\log \; {\rm det} \;
\mu^{-2}\biggr[-\delta_{ab}\cstok{\ }
+{\left. {\partial^{2} V_{0}\over \partial \varphi_{a}
\partial \varphi_{b}} \right|}_{\varphi_{0}}
\biggr]
&(3.3)\cr}
$$
since the ghost determinant cancels the longitudinal one.

To understand how to generalize (3.3) to $SO(10)$ GUT theories,
we have to bear in mind only the first line of (3.3), since,
by virtue of the Coleman-Weinberg mechanism, only gauge-field
loop diagrams contribute to the symmetry-breaking pattern in
the early universe (Allen 1983, Allen 1985, Buccella {\it et
al} 1992). Denoting by $\psi$ the logarithmic derivative
of the $\Gamma$ function, and
defining the functions $\cal A$ and $P$ by means of
$$ \eqalignno{
{\cal A}(z) &\equiv {z^{2}\over 4}+{z\over 3}
-\int_{2}^{{3\over 2}+\sqrt{{1\over 4}-z}}
y\Bigr(y-{3\over 2}\Bigr)(y-3)\psi(y) \; dy \cr
&-\int_{1}^{{3\over 2}-\sqrt{{1\over 4}-z}}
y\Bigr(y-{3\over 2}\Bigr)(y-3) \psi(y) \; dy
&(3.4)\cr}
$$
$$
P(z) \equiv {z^{2}\over 4}+z
\eqno (3.5)
$$
one thus finds for the $SU(5)$ model (Allen 1985)
$$
V(\varphi_{0})=V_{0}(\varphi_{0})-{1\over 2\Omega}
\sum_{l=1}^{24}\Bigr[{\cal A}(a^{2}m_{l}^{2})
+P(a^{2}m_{l}^{2})
\log(\mu^{2}a^{2})\Bigr]
\eqno (3.6)
$$
where the $m_{l}^{2}$ are the 24 eigenvalues of the mass
matrix $M_{ab}^{2}$.

In the case of the $SO(10)$ GUT model, the same method used
for $SU(5)$ in Allen (1985) shows that the one-loop effective
potential $V$ takes the form (see appendix)
$$
V={\widehat V}_{c}-{1\over 2\Omega}\sum_{i=1}^{45}
\Bigr[{\cal A}(a^{2}m_{i}^{2})+P(a^{2}m_{i}^{2})
\log(\mu^{2}a^{2})\Bigr]
\; \; \; \;
\eqno (3.7)
$$
where (cf (2.13))
$$
{\widehat V}_{c}=\biggr({\alpha \over 8}f_{\alpha}
+{\gamma \over 4}f_{\gamma}+{\delta \over 9}f_{\delta}
+(\lambda -\delta)\biggr){\| \phi_{0} \|}^{4}
+{R\over 12}{\| \phi_{0} \|}^{2}.
\eqno (3.8)
$$
The corresponding one-loop
effective potential in (3.7) is obtained by inserting the
following formulae:
$$ \eqalignno{
\sum_{i=1}^{45}{\cal A}(a^{2}m_{i}^{2})&=
6{\cal A}\biggr[a^{2}m^{2}\Bigr({2\over 3}
z_{1}^{2}+{z_{2}^{2}\over 2}
+z_{3}^{2}+{2\sqrt{2}\over 3}z_{1}z_{2}
+\sqrt{{2\over 3}} \; z_{2}z_{3}\Bigr)\biggr]\cr
&+6{\cal A}\biggr[a^{2}m^{2}\Bigr({2\over 3}z_{1}^{2}
+{z_{2}^{2}\over 2}+z_{3}^{2}-\sqrt{{2\over 3}} \;
z_{2}z_{3}\Bigr)\biggr]\cr
&+{\cal A}\biggr[a^{2}m^{2}{z_{2}^{2}\over 2}\biggr]
+3{\cal A}\biggr[a^{2}m^{2}{2\over 3}\Bigr(z_{1}^{2}+z_{2}^{2}
\Bigr)\biggr]
&(3.9)\cr}
$$
$$ \eqalignno{
\sum_{i=1}^{45}P(a^{2}m_{i}^{2})&=
a^{2}m^{2}\biggr(10 z_{1}^{2}+4\sqrt{2} \; z_{1}z_{2}
+{17\over 2}z_{2}^{2}+12z_{3}^{2}\biggr)\cr
&+a^{4}m^{4} \biggr({5\over 3}z_{1}^{4}
+{4\over 3}\sqrt{2}\; z_{1}^{3}z_{2}
+4z_{1}^{2}z_{2}^{2}\cr
&+\sqrt{2} \; z_{1}z_{2}^{3}+{55\over 48}z_{2}^{4}
+{4\over \sqrt{3}}z_{1}z_{2}^{2}z_{3}
+4z_{1}^{2}z_{3}^{2}\cr
&+2\sqrt{2}z_{1}z_{2}z_{3}^{2}+5z_{2}^{2}z_{3}^{2}
+3z_{3}^{4}\biggr).
&(3.10)\cr}
$$
\vskip 0.3cm
\leftline {\bf 4. Flat-space limit}
\vskip 0.3cm
\noindent
The one-loop effective potential (3.7)-(3.10) can hardly be
used for an analytic or numerical study of the absolute
minima, since it involves a large number of complicated
contributions. We therefore begin by studying its flat-space
limit, i.e. its asymptotic behaviour when the 4-sphere radius
$a$ tends to $\infty$. The corresponding asymptotic form of
${\cal A}(z)$ is (Allen 1985)
$$
{\cal A}(z) \sim -\biggr({z^{2}\over 4}+z+{19\over 30}\biggr)
\log(z)+{3\over 8}z^{2}+z+{\rm const.}+{\rm O}(z^{-1}).
\eqno (4.1)
$$
For the purpose of numerical analysis at $a \rightarrow
\infty$, the expansion (4.1) can be further approximated
as
$$
{\cal A}(z) \sim {z^{2}\over 8}\Bigr(3-\log(z^{2})\Bigr).
\eqno (4.2)
$$
Thus, defining (cf end of section 2)
$$
h_{1} \equiv {2\over 3}z_{1}^{2}+{z_{2}^{2}\over 2}
+z_{3}^{2}+{2\sqrt{2}\over 3}z_{1}z_{2}
+\sqrt{{2\over 3}} \; z_{2}z_{3}
\eqno (4.3)
$$
$$
h_{2} \equiv {2\over 3}z_{1}^{2}+{z_{2}^{2}\over 2}
+z_{3}^{2}-\sqrt{{2\over 3}} \; z_{2}z_{3}
\eqno (4.4)
$$
$$
h_{3} \equiv {z_{2}^{2}\over 2}
\eqno (4.5)
$$
$$
h_{4} \equiv {2\over 3}\Bigr(z_{1}^{2}+z_{2}^{2}\Bigr)
\eqno (4.6)
$$
$$
h_{5}^{2} \equiv {3\over 2}h_{1}^{2}+{3\over 2}h_{2}^{2}
+{h_{3}^{2}\over 4}+{3\over 4}h_{4}^{2}
\eqno (4.7)
$$
$$
y \equiv {m\over \mu}
\eqno (4.8)
$$
equations (3.7)-(3.10) and (4.2) lead to
$$ \eqalignno{
{V\over \mu^{4}} & \sim {{\widehat V}\over \mu^{4}}
-{3\over 8\pi^{2}}y^{4}
\biggr[{3\over 4}h_{1}^{2}\Bigr(3-\log(h_{1}^{2})\Bigr)
+{3\over 4}h_{2}^{2}\Bigr(3-\log(h_{2}^{2})\Bigr) \cr
&+{h_{3}^{2}\over 8}\Bigr(3-\log(h_{3}^{2})\Bigr)
+{3\over 8}h_{4}^{2}\Bigr(3-\log(h_{4}^{2})\Bigr)
-h_{5}^{2}\log\Bigr(y^{2}\Bigr)\biggr].
&(4.9)\cr}
$$
The problem now arises to find the absolute minima of the
potential (4.9) by numerical methods. Since $z_{1},
z_{2}, z_{3}$ lie on a unit 2-sphere, they can be
expressed as $z_{1}=\sin(\theta)\cos(\varphi),
z_{2}=\sin(\theta)\sin(\varphi),z_{3}=\cos(\theta)$.
For given values of
the parameters $\alpha, \gamma, \lambda, \delta$ appearing
in (3.8), we have thus to minimize with respect to
$\theta, \varphi, y$. For this purpose, we point out that
$y_{{\rm min}}$ should be $\leq 1$, since it is
the ratio of the gauge-boson mass to the cut-off value.
Hence one gets a further restriction on $\lambda$ which,
combined with the inequality (2.20), yields the
sufficient condition
$$
\lambda \geq \lambda_{0}+{\mid \alpha \mid \over 8}
\Bigr(f_{\alpha}\Bigr)_{{\rm max}}
+{\mid \delta \mid \over 9}
\Bigr(f_{\delta}\Bigr)_{{\rm max}}.
\eqno (4.10)
$$
With our notation, $\lambda_{0}$ is given by
$$ \eqalignno{
\lambda_{0} & \equiv {3{\cal G}^{4}\over 8\pi^{2}}
\biggr[{3\over 4}h_{1}^{2}\Bigr(3-\log(h_{1}^{2})\Bigr)
+{3\over 4}h_{2}^{2}\Bigr(3-\log(h_{2}^{2})\Bigr) \cr
&+{h_{3}^{2}\over 8}\Bigr(3-\log(h_{3}^{2})\Bigr)
+{3\over 8}h_{4}^{2}\Bigr(3-\log(h_{4}^{2})\Bigr)
-{1\over 2}h_{5}^{2}\log\Bigr(y^{2}\Bigr)
\biggr]_{{\rm min}(\theta,\varphi)}
-{\widetilde f}_{{\rm min}}
\; \; \; \;
&(4.11)\cr}
$$
where ${\widetilde f}$ is the function such that
$\Bigr({\widetilde f}+\lambda \Bigr)y^{4}={\widehat V}$.
The corresponding numerical analysis, carried out by using
the MINUIT minimization program available in the CERN libraries,
shows that the absolute minimum always lies in the
$\theta=0$ direction. This is the $SU(4)_{PS} \otimes
SU(2)_{L} \otimes SU(2)_{R}$ symmetry-breaking direction
(see (2.12c)). Thus, as far as the
absolute-minimum direction is concerned,
the flat-space limit of the one-loop
calculation about a de Sitter background does not change the
results found in Buccella {\it et al} (1986), where the tree-level
potential in flat space-time was studied. Remarkably, since the value
of $y$ leading to the absolute minimum of $V$
in the presence of symmetry breaking has been
found to be $y_{{\rm min}} \in [0.4,0.8]$ in the regions where
the inequality (4.10) is satisfied, one can evaluate $\mu$
from (4.8) as
$$
\mu={M_{X}\over y_{{\rm min}}}.
\eqno (4.12)
$$
This formula for $\mu$ can be used to derive the values of
the 4-sphere radius corresponding to given values of
the dimensionless parameter $\mu a$ (see below).

To complete this section, we find it helpful for the reader
to evaluate the behaviour of the flat-space-limit one-loop
effective potential as $\theta \rightarrow 0$, since $\theta=0$
yields the absolute-minimum direction as we just said.
The analytic calculation shows that the potential
$V$ in (4.9) obeys the relation
$$
\lim_{\theta \rightarrow 0}
V(\lambda,y,\theta) \equiv V_{{\rm lim}}=
{y^{4}\over \pi^{2}} \left[{625 \over 4}\lambda
-{27\over 16} + {9\over 4} \log(y) \right].
\eqno (4.13)
$$
The corresponding behaviours of $V_{{\rm lim}}$ for various values
of $\lambda$ ($\lambda=0.03,0.02,0.015,0.012$)
are plotted in figure 1, when $y \in [0,1]$.
\vskip 0.3cm
\leftline {\bf 5. Numerical evaluation of the absolute minima}
\vskip 0.3cm
\noindent
As one can see from equations (3.7)-(3.10), the one-loop
effective potential for our cosmological model takes a
complicated form, and it is not clear whether curvature can
modify the results of section 4, once the same values for
$\alpha,\gamma,\delta,\lambda$ have been chosen. The
corresponding absolute minima have been evaluated using again
the MINUIT minimization program
and choosing different values for the dimensionless parameter
$\mu a$, since it is convenient to work with
the dimensionless form of the one-loop potential,
obtained dividing (3.7)-(3.10)
by $\mu^{4}$. Of course, the parameters
in the effective potential are
$\alpha,\gamma,\delta,\lambda,\mu a$, whereas the arguments are
$y,\theta,\varphi$.

Interestingly, if $\mu a$ is $\leq 1$, the term
${R\over 12}{\| \phi \|}^{2}$ in the potential (3.7)-(3.10)
dominates over all other contributions, and hence does not
lead to any symmetry breaking. Thus, only intermediate values
of $\mu a$ are relevant for the symmetry-breaking pattern. In
this case the absolute minima are still found to be
$SO(6) \otimes SO(4)$-invariant (for them $\theta=0$), providing
the inequality (4.10) is satisfied. In figures 2-3, obtained
setting $\mu a=30,300$ respectively, the one-loop effective
potential is plotted as a function of $y$ when $\alpha=\gamma
=\delta=0$. Note that these particular values are chosen since
the flat-space effective potential (4.13) is independent of
$\alpha,\gamma,\delta$. Hence $\alpha,\gamma,\delta$ can be
set to zero for simplicity when curvature vanishes, whereas in
the presence of curvature they are set to zero to compare the
flat-space analysis with the de Sitter case.

A naturally occurring question is what can be learned from
the comparison of figure 1 with figures 2-3. Indeed, the
values of the independent variable $y$ for which the absolute
minima are attained are modified in the presence of
curvature. The smaller $\mu a$ (stronger curvature),
the more substantial the change of the shape of our
plots. In particular, figure 2 shows that for $\mu a=30$
and $\alpha=\gamma=\delta=0$ no symmetry breaking occurs
even though for other choices of values for $\alpha,\gamma,
\delta$, non-trivial absolute minima are present.
By contrast, from figure 3, corresponding to $\mu a=300$,
the effects of curvature on the absolute minima of the
potential can be easily seen (cf figure 1).

Remarkably, our numerical investigation shows that, providing the
mass matrix is positive-definite, the potential is bounded
from below, and the gauge-boson mass remains smaller than
the cut-off value, the absolute-minimum direction remains
$SO(6) \otimes SO(4)$-invariant in flat or de Sitter space,
if the tree-level potential has this invariance property.
To help the reader, table III shows, for the same values
of the parameters used in figure 3, the values taken by
$y_{{\rm min}}$ and the corresponding values of the
dimensionless one-loop effective potential. The $\theta$
and $\varphi$ entries are omitted since $\theta=0$ and
$\varphi$ is undetermined in the presence of spontaneous
symmetry breaking along the $SO(6) \otimes SO(4)$ direction.
\vskip 0.3cm
\leftline {\bf 6. Results and concluding remarks}
\vskip 0.3cm
\noindent
The main results of our investigation are as follows.

First, the one-loop effective potential of $SO(10)$ GUT
theories in de Sitter space has been obtained for the first
time. This analytic result represents the continuation of
the program initiated in Allen (1985), where the tools
necessary for any non-Abelian gauge theory in de Sitter
space were described in detail. Note that, while (3.7)
holds for any irreducible representation of $SO(10)$, (3.8)
relies on the $\underline {210}$ representation, and
(3.9)-(3.10) lead to a {\it particular} form of such
potential, once $SU(3) \otimes SU(2) \otimes U(1)$
invariance for the mass matrix is required to agree with
electroweak symmetry.

Second, the flat-space limit of the corresponding
Coleman-Weinberg effective potential has been evaluated
for the $\underline {210}$ representation.

Third, the numerical analysis of absolute minima has been
carried out in the case of the mass matrix relevant for
low-energy-limit phenomenology. Interestingly, de Sitter curvature
does not affect the flat-space symmetry-breaking
pattern, leading only to the
$SO(6) \otimes SO(4)$ symmetry-breaking direction.

A naturally occurring question is whether the analytic study
of absolute minima can be performed, to check the results
of our numerical investigation. In principle, this research
appears possible, although it goes beyond the author's
computational skills, due to the many parameters appearing
in the $SO(10)$ effective potential. For the time being,
we should emphasize that our results, although obtained after
a time-consuming numerical analysis, remain preliminary.

It has been our task to work under the restrictive
conditions summarized at the end of section 5, while
other forms of the mass matrix remain unknown in the
literature. Thus, a complete mathematical treatment similar
to what was done in Allen (1985) for $SU(5)$ theories is
lacking, and appears to be a topic for further research.
Moreover, since the Higgs field (if it exists) is actually
varying in time, it appears necessary to evaluate the
one-loop effective potential of non-Abelian gauge theories
in closed FRW cosmologies, de Sitter being just a particular
case. This more complicated analysis would supersede the
approximations made in Allen (1983) and Esposito {\it et al}
(1993) to study the evolution of the early universe.
\vskip 0.3cm
\leftline {\bf Appendix}
\vskip 0.3cm
\noindent
To obtain the one-loop effective potential
(3.7)-(3.8) one starts
from the bare, Euclidean-time Lagrangian (cf (3.1))
$$
L={1\over 4}{\rm Tr}\Bigr({\bf F}_{\mu \nu}{\bf F}^{\mu \nu}\Bigr)
+{1\over 2}{\rm Tr}\Bigr(D_{\mu}\phi \Bigr)\Bigr(D^{\mu}\phi \Bigr)
+V_{0}(\phi)
\eqno (A.1)
$$
where $D_{\mu} \equiv \partial_{\mu}
-i{\cal G} A_{\; \; \mu}^{ab}
\; T^{ab} \; \forall a,b=0,...,9$. According to the
background-field method, one expands the field $\phi$ as
$$
\phi=\phi_{0}+{\widetilde \phi}
\eqno (A.2)
$$
where $\phi_{0}$ is the background value, and
$\widetilde \phi$ is a perturbation. The 4-metric $g$ is
also expanded as in Allen (1983, 1985). The resulting one-loop
form of $L$, i.e. the Lagrangian quadratic in the perturbations,
is
$$ \eqalignno{
L^{(1)}&={1\over 4}{\rm Tr} \Bigr({\bf F}_{\mu \nu}
{\bf F}^{\mu \nu}\Bigr) +{1\over 2}
{\rm Tr} \Bigr(D_{\mu}{\widetilde \phi}\Bigr)
\Bigr(D^{\mu}{\widetilde \phi}\Bigr) \cr
&+{i{\cal G} \over 2}\biggr[{\Bigr(\nabla_{\mu}A^{\mu}\Bigr)}^{lm}\biggr]
\biggr[<{\widetilde \phi} \mid T^{lm} \mid \phi_{0}>
-<\phi_{0} \mid T^{lm} \mid {\widetilde \phi}> \biggr] \cr
&+{{\cal G}^{2}\over 2}A_{\mu}^{lm}
<\phi_{0} \mid T^{lm}T^{pq} \mid \phi_{0}>
A^{pq \; \mu} +V_{0}
+{1\over 2}{\Bigr({\widetilde \phi}\Bigr)}^{2}
{\left. {\partial^{2}V_{0} \over \partial \phi^{2}}
\right|}_{\phi=\phi_{0}}.
&(A.3)\cr}
$$
Moreover, the gauge-averaging term we are looking for
is (cf Allen 1985)
$$
L_{\rm gauge}={{\widetilde \alpha}\over 2}
{\rm Tr}{\biggr[{\Bigr(\nabla_{\mu}A^{\mu}\Bigr)}^{lm}
+{\widetilde \beta}
\Bigr(<{\widetilde \phi} \mid T^{lm} \mid \phi_{0}>
-<\phi_{0} \mid T^{lm} \mid {\widetilde \phi}>\Bigr)
\biggr]}^{2}.
\eqno (A.4)
$$
By virtue of equations (A.3)-(A.4), cross-terms disappear
in $L^{(1)}+L_{\rm gauge}$ if and only if
${\widetilde \beta}=-{i{\cal G} \over 2}{\widetilde \alpha}^{-1}$.
This leads to
$$ \eqalignno{
L^{(1)}+L_{\rm gauge}&
={1\over 4}{\rm Tr} \Bigr({\bf F}_{\mu \nu}
{\bf F}^{\mu \nu}\Bigr)+{1\over 2}{\rm Tr} \Bigr(D_{\mu}
{\widetilde \phi}\Bigr)
\Bigr(D^{\mu}{\widetilde \phi}\Bigr) \cr
&+{{\widetilde \alpha}\over 2}{\rm Tr}
{\biggr[{\Bigr(\nabla_{\mu}A^{\mu}\Bigr)}^{lm}\biggr]}^{2}
-{{\cal G}^{2}\over 8 {\widetilde \alpha}}{\rm Tr}
{\biggr[<{\widetilde \phi} \mid T^{lm} \mid \phi_{0}>
-<\phi_{0} \mid T^{lm} \mid {\widetilde \phi}> \biggr]}^{2}\cr
&+{{\cal G}^{2}\over 2}A_{\mu}^{lm} <\phi_{0} \mid T^{lm}T^{pq} \mid
\phi_{0}>A^{pq \; \mu} +V_{0}
+{1\over 2}{\Bigr({\widetilde \phi}\Bigr)}^{2}
{\left. {\partial^{2}V_{0} \over \partial \phi^{2}}
\right|}_{\phi=\phi_{0}}.
&(A.5)\cr}
$$
By splitting the gauge potential into transverse and longitudinal
part on the $S^{4}$ background, and following Allen (1985), one
obtains an equation similar to (3.3), where the mass matrix has
45 eigenvalues rather than 24. Hence (3.7) is proved.
\vskip 5cm
\leftline {\bf Acknowledgments}
\vskip 0.3cm
\noindent
We are much indebted to Franco Buccella for teaching us
all what we know about $SO(10)$ GUT theories and for
reading the manuscript, and to Ofelia Pisanti for providing
the ranges of the parameters studied in section 2.
Anonymous referees made comments which led to a substantial
improvement of the original manuscript.
This work was supported in part by DOE and by the NASA
(NAGW-2381) at Fermilab.
\vskip 0.3cm
\leftline {\bf References}
\vskip 0.3cm
\parindent=0pt
\everypar{\hangindent=20pt \hangafter=1}

Acampora F, Amelino Camelia G, Buccella F, Pisanti O,
Rosa L and Tuzi T 1994 {\it Proton decay and neutrino
masses in SO(10)} (DSF preprint 93/52, submitted to
{\it Rev. Mod. Phys.})

Allen B 1983 {\it Nucl. Phys.} B {\bf 226} 228

Allen B 1985 {\it Ann. Phys., N.Y.} {\bf 161} 152

Buccella F, Cocco L, Sciarrino A and Tuzi T 1986
{\it Nucl. Phys.} B {\bf 274} 559

Buccella F 1988 Spontaneous symmetry breaking in unified
theories with gauge group $SO(10)$ {\it Lochau 1988,
Proceedings, Symmetries in Science III} ed B Gruber
and F Iachello

Buccella F, Esposito G and Miele G 1992 {\it Class. Quantum
Grav.} {\bf 9} 1499

Chang D, Mohapatra R N and Parida M K 1984 {\it Phys. Rev. Lett.}
{\bf 52} 1072

Coleman S and Weinberg E 1973 {\it Phys. Rev.} D {\bf 7} 1888

Cornwell J F 1984a {\it Group Theory in Physics, Vol. I}
(New York: Academic Press)

Cornwell J F 1984b {\it Group Theory in Physics, Vol. II}
(New York: Academic Press)

Esposito G, Miele G and Rosa L 1993 {\it Class. Quantum Grav.}
{\bf 10} 1285

Esposito G 1994 {\it Quantum Gravity,
Quantum Cosmology and Lorentzian
Geometries} Lecture Notes in Physics, New Series m:
Monographs vol m12 second corrected and enlarged edn
(Berlin: Springer)

Fritzsch H and Minkowski P 1975 {\it Ann. Phys., N.Y.}
{\bf 93} 193

Hawking S W 1977 {\it Commun. Math. Phys.} {\bf 55} 133

Kuzmin V and Shaposhnikov N 1980 {\it Phys. Lett.}
{\bf 92B} 115

O'Raifeartaigh L 1986 {\it Group Structure of Gauge Theories}
(Cambridge: Cambridge University Press)

Pati J C and Salam A 1973 {\it Phys. Rev.} D {\bf 8} 1240

Tuan S F 1992 {\it Mod. Phys. Lett.} A {\bf 7} 641
\vskip 0.3cm
\leftline {Figure captions:}
\vskip 0.3cm
\noindent
{\bf Figure 1.} Flat-space limit of the dimensionless
one-loop effective potential at $\theta=0$ (4.13) versus
$y$ (4.8) is here shown. The curves correspond to the
$\lambda$-values 0.03, 0.02, 0.015, 0.012 respectively.
\vskip 0.3cm
\noindent
{\bf Figure 2.} The dimensionless form of the one-loop
effective potential in de Sitter space at $\theta=0$,
and $\mu a=30$ versus $y$ is here shown. The curves
correspond to $\alpha,\gamma,\delta=0$ and the same values
of $\lambda$ of figure 1.
\vskip 0.3cm
\noindent
{\bf Figure 3.} The one-loop potential of figure 2 is
evaluated for $\mu a=300$.
\vskip 0.3cm
\noindent
Table captions:
\vskip 0.3cm
{\bf Table I.} The intermediate symmetries, the Higgs directions
and the IRR's of $SO(10)$ used for the Higgs scalar fields are
here reported for the most physically relevant $SO(10)$ GUT
models (Acampora {\it et al} 1994). With our notation,
$\omega_{ab}$ denotes the 54-dimensional irreducible
representation of $SO(10)$.
\vskip 0.3cm
\noindent
{\bf Table II.} For the same models of table I, the masses of gauge
bosons are shown, following Acampora {\it et al} (1994).
\vskip 0.3cm
\noindent
{\bf Table III.} For the same values of the parameters used in
figure 3, the values taken by $y_{{\rm min}}$ and by the
dimensionless one-loop effective potential are shown.

\bye